\documentclass[manuscript]{aastex}

\shorttitle{N isotopic ratios in carbon stars}
\shortauthors{Hedrosa et al.}

\begin{document}


\title{Nitrogen isotopes in AGB carbon stars and presolar SiC grains:\\ 
a challenge for stellar nucleosynthesis}


\author{R. P. Hedrosa\altaffilmark{1}, C. Abia\altaffilmark{1}}
\affil{Departamento de F\'\i sica Te\'orica y del Cosmos, Universidad de 
Granada, 18071 Granada, Spain}

\author{M. Busso\altaffilmark{2,3}}
\affil{Dipartimento di Fisica, Universit\`{a} di Perugia, 06123 Perugia, Italy}
\affil{INFN, Sezione di Perugia, 06123 Perugia, Italy}

\author{S. Cristallo\altaffilmark{4}}
\affil{INAF, Osservatorio di Collurania, 64100 Teramo, Italy}

\author{I. Dom\'\i nguez\altaffilmark{1}, S. Palmerini\altaffilmark{1}}
\affil{Departamento de F\'\i sica Te\'orica y del Cosmos, Universidad de 
Granada, 18071 Granada, Spain}

\author{B. Plez\altaffilmark{5}}
\affil{Laboratoire Univers et Particules de Montpellier, Universit\'{e} 
Montpellier II, CNRS, 34095 Montpellier, France }

\and

\author{O. Straniero\altaffilmark{4}}
\affil{INAF, Osservatorio di Collurania, 64100 Teramo, Italy}

\begin{abstract}
Isotopic ratios of C, N, Si, and trace heavy elements in presolar SiC grains from meteorites
provide crucial constraints to nucleosynthesis. A long-debated issue is the origin of
the so-called A+B grains, as for them no stellar progenitor has so far been clearly 
identified on observational grounds. We report the first spectroscopic measurements 
of $^{14}$N/$^{15}$N ratios in Galactic carbon stars of different spectral types and show 
that J- and some SC-type stars might produce A+B grains, even for $^{15}$N 
enrichments previously attributed to novae. We also show that most mainstream (MS) grains are 
compatible with the composition of N-type stars, but might also descend, in some cases, 
from SC stars. From the theoretical point of view, no astrophysical scenario can 
explain the C and N isotopic ratios of SC, J and N-type carbon stars together, as well as 
those of many grains produced by them. This poses urgent questions to stellar physics.
\end{abstract}


\section{Introduction}
After the exhaustion of He in the core, stars of mass 
$0.8 \lesssim M/M_{\odot} \lesssim 8$ become very luminous and cool and climb the so
called Asymptotic Giant Branch (AGB). AGB stars
are powered by two nuclear shells, burning H and He alternatively; He, in
particular, burns recurrently in short explosive events called 
thermal pulses. After most of them, the convective envelope penetrates downward 
bringing to the surface H- and He-burning products in a
phenomenon called third dredge up (TDU). Carbon is the main product 
of He burning, thus AGB stars, from their initially O-rich composition, get enriched 
in carbon (and in other He-burning products, like s-elements). For suitable values of the
envelope mass (eroded by mass loss) AGB stars finally achieve an
abundance ratio C/O $> 1$ (by number), in which case they become
carbon stars of a class called C(N) (or N-type). This typically occurs between 1.5 
and 3-4 M$_{\odot}$ for solar chemical composition \citep{ab1,cris}. 
Above M $\gtrsim 4$ M$_{\odot}$ stars cannot become carbon rich, both because of the 
large envelope mass to pollute, and because the hot temperature at its base induces 
CN-cycling, burning the dredged-up carbon in a process called hot bottom burning, or HBB 
\citep{ren}.

It was inferred spectroscopically that composition
changes induce a spectral type evolution along the sequence \citep{wal}:
M$\rightarrow$MS$\rightarrow$S$\rightarrow$SC$\rightarrow$C(N) although there are some
doubts on the nature of SC stars, see below. The composition determines the type of condensates 
forming in the circumstellar envelope. Oxides and silicates form in O-rich AGB stars (M, MS, S)                
while C-rich stars are parents to SiC and graphite dust. After its ejection into 
the interstellar medium by stellar winds, this cosmic dust can be trapped in meteorites, 
that are now recovered in the Solar System. Among these dust particles, SiC grains are probably the best studied ones
(see e.g. Davis 2011). They are classified on the basis of
their nitrogen, carbon and silicon isotopic ratios \citep[see e.g.,][]{nit}. The so-called
mainstream (MS) grains which constitute 93\% of all SiC grains, show a huge range in the $^{14}$N/$^{15}$N 
ratio (from 10 to 10000). Moreover, these grains show isotopic anomalies that can be only 
explained if they have been formed from material exposed to s-process nucleosynthesis,
which is thought to occur during the AGB phase of low-mass stars \citep[see e.g.,][]{stra}. 
On the other hand, grains of type A+B ($\sim$ 4\% of presolar SiC) show also a large spread
in the N ratio, but low $^{12}$C/$^{13}$C ratios ($< 10$); their origin is still unclear \citep{ama}.

Spectroscopically, C(N) stars show strong CN and C$_2$ bands. They also display
absorption lines of F and s-elements, whose enrichment is in 
good agreement with stellar and nucleosynthesis models \citep{ab1,ab2,cris},
so that the AGB evolutionary stages are thought to be well understood.
Their $^{12}$C/$^{13}$C ratios are typically $\gtrsim30$, averaging 
at $\sim 60$. Other types of carbon-rich giants exist, but of a more unclear origin. 
Among them, SC-type stars show molecular bands indicating C/O ratios close to unity; s-element 
enhancements are not always present and the $^{12}$C/$^{13}$C ratios range from CN-cycle equilibrium 
(3$-$4) up to $\sim 100$. A few SC stars are super Li-rich \citep{ab3}, with Li abundances 
larger by 4-5 orders of magnitude than for C(N) giants. They also show the largest 
F enhancements\footnote{Here we adopt the definition [A/B]$=$log$(N_A/N_B)-$log$(N_A/N_B)_{\odot}$
where log ($N_A/N_H$) is the abundance by number of the element $A$.}, [F/Fe] $\sim 1$ \citep{ab2}, in
solar metallicity ([Fe/H] $\sim 0$)  carbon stars. Furthermore, they seem to be
on average more luminous than C(N) giants \citep{gua}, suggesting stellar masses
$\gtrsim$ 4 M$_{\odot}$, thus casting doubts on their position in the spectral
sequence going from M to C(N) types. Another subgroup of carbon stars, the J-type,
shows strong features of $^{13}$C-bearing molecules, indicating $^{12}$C/$^{13}$C $ \lesssim
15$. They have no s-element enhancement and are, in most cases ($\sim
80\%$), moderately Li-rich \citep{ab3}. Their relation to the quoted
spectral sequence is unclear. Several J stars ($\sim 30\%$) show infrared emission
lines, associated with silicate dust: a peculiar property, given their C-rich composition, 
although chemical kinetics allows for the formation of O-based compounds in C-rich
environments \citep{codust}. The emission seems in any case to come from O-rich
discs in binary systems \citep{chi}, indicating binarity as a common property of these stars.

Here we derive for the first time the $^{14}$N/$^{15}$N ratios in a sample of near solar metallicity
Galactic carbon stars of different spectral types. Our analysis reveals that the N
isotopic ratios in N-type stars cover nicely the range found in the mainstream SiC grains,
while those derived in J- and SC-type support an origin for A+B grains in these 
peculiar stars. We discuss briefly the results in the framework of the standard AGB phase
stellar evolution and conclude that no known evolutionary scenario can explain
the full range of $^{14}$N/$^{15}$N ratios found in these stars.

\section{Observations and analysis}
We obtained very high-resolution echelle spectra ($R\approx 170000$)
of nineteen N-, eight J- and eight SC-type Galactic carbon stars of near solar metallicity with
the SARG spectrograph at the 3.5m TNG telescope. The signal-to-noise ratio in the spectral
region of interest ($\sim 8000$ {\AA}) was typically $\gtrsim 300$.
In this wavelength interval there are various $^{12}$C$^{15}$N absorption
features sensitive to the $^{15}$N abundance. The CN line list by \citet{hill} was improved to 
allow the identification of $^{12}$C$^{14}$N, $^{13}$C$^{14}$N, and $^{12}$C$^{15}$N lines. 
First, wavelengths were improved using the energy levels of Ram, Wallace, and Bernath (2010) 
for $^{12}$C$^{14}$N, and of \citet{ram} for $^{13}$C$^{14}$N. They were supplemented by 
wavelengths from \citet{kot} or calculated by extrapolation of the 
molecular constants when needed. Isotopic shifts were computed for all isotopic combinations, 
using the usual isotope relationship for the Dunham coefficients \citep{tow}. 
It appeared that $^{13}$C$^{14}$N line positions computed in this way were systematically displaced 
relative to the laboratory data by \citet{ram}. We therefore anticipated that the line positions  
for other isotopologues would be shifted as well. 
We realized  that the ratio of the computed isotopic 
shifts for 2 different isotopologues were close to a 
constant,  i.e. $C_{25/34}\approx 0.73${\footnote{Isotopic shifts are calculated 
relative to the majority species ${\rm ^{12}C^{14}N}$. $C_{25/34}=(\lambda_{25}-\lambda_{24})/(\lambda_{34}-\lambda_{24})$ 
is the mean ratio of the isotopic shifts of lines of the 
${\rm ^{12}C^{15}N}$ and ${\rm ^{13}C^{14}N}$  isotopologues, for a given band.}
in the case of $^{12}$C$^{15}$N and $^{13}$C$^{14}$N in the spectral region of interest. 
We therefore computed a correction to our calculated 
$^{12}$C$^{15}$N isotopic shift, scaling the difference for $^{13}$C$^{14}$N between 
Ram et al.'s measurement  and our calculated value of the isotopic shift with the $C_{25/34}$ factor.
 We could compare this calculation to observed $^{12}$C$^{15}$N line positions kindly sent to us by R. Colin. 
The comparison showed a good agreement, sufficient to give us confidence in our identifications 
of isotopic lines in stellar spectra. The used features were carefully selected to avoid blending and 
excluding lines for which, the pseudo-continuum position was uncertain. Moreover, only
$^{12}$C$^{15}$N features in the linear part of the curve-of-growth were used.
This resulted in a few useful lines located near 7980, 7985, 8030, 8038, 
and 8064 {\AA} (see Table~\ref{tab1}). For other molecular and atomic lines contributing in
this region we used line lists from previous works \citep{ab1,ab2}.
Stellar parameters (T$_{eff}$, gravity, metallicities, C/O and
$^{12}$C/$^{13}$C ratio) were taken from the literature (see references in Table~\ref{tab2}). For some stars (see
Table~\ref{tab2}), carbon and oxygen abundances were derived from a few weak, unblended C$_2$ and
CO lines in the 2.2 $\mu$m region using high signal-to-noise and resolution (R$\sim= 65000$)
spectra (kindly provided by K. Hinkle) obtained at the 4 m Kit Peak Observatory telescope
using a Fourier transform spectrograph. Then, the N abundance as well as the final C/O ratio
were derived from CN lines in the 8000 {\AA} region in an iterative way until agreement with
the values obtained in the infrared spectral range was reached. We note however, that 
uncertainties in the absolute abundance of N within $\pm 0.3$ dex does not affect the 
nitrogen ratio derived.


A C-rich spherical MARCS \citep{gus} model atmosphere was chosen for each star according to its stellar
parameters, and synthetic LTE spectra were calculated in the 8000 {\AA}
region, using the Turbospectrum v10.1 code \citep{plez}. Theoretical spectra were convolved with a Gaussian 
function with the corresponding FWHM to mimic the spectral resolution in each range plus the 
macroturbulence parameter (9-13 km s$^{-1}$). We used $\chi^2$ {\bf minimization techniques}
to determine the $^{14}$N/$^{15}$N ratios providing the
best fit to each $^{12}$C$^{15}$N feature. The goal was to fit not
only the selected lines, but also the overall shape of the
spectra. The N isotopic ratios thus derived were then combined to obtain an
average. The N ratios obtained from the $^{12}$C$^{15}$N features at $\lambda 7980$, and 
$\lambda 8064$ {\AA} were considered twice in deriving this average. These features 
are the most sensitive to $^{14}$N/$^{15}$N variations. In this way we measured reliable N isotopic 
ratios for 22 stars of our
sample; in a few cases we did not detect $^{15}$N and for the rest we
established lower limits on $^{14}$N/$^{15}$N (see Table~\ref{tab2}). In most cases, the
overall uncertainty in the N ratios is estimated to be less than a {\bf factor of four}. 
This mainly reflects the sensitivity to changes in the atmospheric parameters 
adopted, plus the dispersion in the $^{14}$N/$^{15}$N ratio derived among the 
different features. This also includes the uncertainty in the placement
of the spectral continuum ($\leq 3\%$) and in the calculated
wavelength of the $^{12}$C$^{15}$N features ($\leq 15$ m{\AA}).
Thus, to minimize the errors we performed a relative line-by-line analysis with
respect to the C(N) star LQ Cyg ([$^{14}$N/$^{15}$N]$_{LQCyg}$) for which we measured $^{14}$N/$^{15}$N$=1170$. 
For this star we obtained a very good global fit to its spectrum. The relative analysis reduced
the dispersion in [$^{14}$N/$^{15}$N]$_{LQCyg}$ derived for a given star (see Table~\ref{tab2}).
We estimate a total uncertainty of $\pm 0.4$ dex for [$^{14}$N/$^{15}$N]$_{LQCyg}$.

Figure 1 shows examples of synthetic fits to 
$^{12}$C$^{15}$N features in four of the studied stars with different
$^{14}$N/$^{15}$N ratios. Figure 2 shows the $^{14}$N/$^{15}$N ratios derived for our sample
stars (normalized to LQ Cyg) versus their $^{12}$C/$^{13}$C ratios. Over-plotted (gray symbols) are the
isotopic ratios measured in MS and A+B SiC grains \citep{hop, ama, hyn}. 
The black point indicates model N and C isotopic ratios for a
2 $M_{\odot}$ AGB star of solar metallicity {\bf at the time that it becomes} C-rich ([$^{14}$N/$^{15}$N]$_{LQCyg}\approx +0.2$, 
$^{12}$C/$^{13}$C$\approx 70$). These values derive from the combined action of the first dredge-up (RGB phase), 
where the $^{14}$N/$^{15}$N ratio grows from the initial (solar) value, 470 \citep{mar} to $\sim 1000$, and the 
subsequent evolution before the C-rich AGB phase. This includes some (small) contribution from non-convective (extra) mixing 
during the RGB phase, as required by observations \citep{pal}. This point, plotted for C/O$=1$, represents a 
{\it lower limit} of N isotopic ratios for solar metallicity C(N) stars. This limit slightly increases 
for increasing stellar mass and decreasing metallicity. Further extra-mixing during the AGB phase would move the point
along the diagonal arrow, while more TDU episodes would increase the $^{12}$C/$^{13}$C
ratio along the right-hand arrow (see Fig.~\ref{fig2}).

\section{Results and discussion}
Among the carbon stars studied, J-type giants are defined mainly by their low
$^{12}$C/$^{13}$C and by the absence of s-elements. Their spectra are difficult to analyze, 
showing broad lines and unidentified features that cannot be well reproduced. In spite of 
this, their C and N isotope ratios closely match those of the A+B grains (see Fig.~\ref{fig2}). 
The observational uncertainties are large, but not enough to hamper this conclusion. This is 
therefore the first experimental, unambiguous evidence ascribing at least part of A+B grains 
to J stars, confirming previous qualitative hints \citep{ama}. Interestingly enough, the fraction 
of A+B grains within all SiC grains ($\sim 5\%$) \citep{dav} is very similar to that of J-type stars among 
all Galactic AGB carbon stars ($\sim 4-10\%$) \citep{bof,bar}.
We also identified, for the first time, a few (although two are lower limits) $^{15}$N-rich
($^{14}$N/$^{15}$N$~\lesssim 1000$, or [$^{14}$N/$^{15}$N]$_{LQCyg}~\lesssim -0.07$) J stars (Fig.~\ref{fig2}): 
an amazing result, with no explanation in red giant models, 
which invariably {\bf predict} $^{14}$N-rich envelopes. Notice that also the very low $^{12}$C/$^{13}$C ratios ($\lesssim 4$), 
shared by many A+B grains and by some J stars, cannot be achieved by nucleosynthesis scenarios 
for red giants except in the case of HBB or extreme extra-mixing processes (see below). These however imply large 
$^{14}$N production and O-rich environments. Low C and N isotopic ratios were so far obtained only in simulations
of nova explosions \citep{jos}; but apart from the fact that novae do not account for the entire 
range of C and N ratios of A+B grains\footnote{Note nevertheless that Nittler \& Hoppe (2005) reported
a SiC grain with supernova isotope signatures but also low C and N ratios.}, there is no known connection 
between novae and J stars}. Actually, until now very few carbon-rich grains can be really ascribed to novae \citep{geh}. 

On the other hand, we notice that almost all the data for C(N) stars lay above 
$^{14}$N/$^{15}$N$\gtrsim 1000$ (or [$^{14}$N/$^{15}$N]$_{LQCyg}\gtrsim-0.07$, Fig. 2) occupying the same
region as many MS grains. As our detection limit is [$^{14}$N/$^{15}$N]$_{LQCyg}\lesssim +0.7$ (or 
$^{14}$N/$^{15}$N$\lesssim 5000$), we cannot 
even exclude the existence of C(N) stars with higher $^{14}$N-enrichments,
as shown by several MS grains. Since MS grains also show s-process 
signatures \citep{zin,gal}, they are believed to form in N-type stars. 
Data of C(N) stars confirm the large spread of N and C isotopic ratios measured in SiC grains, 
stressing our incapability to explain isotopic abundances for several grains and a few C(N) giants with 
any theoretical scenario proposed to date. For instance, the occurrence of any non-convective (extra) mixing 
episode, linking the envelope to regions where proton captures occur, would further increase 
$^{14}$N/$^{15}$N up to values around 10$^4$, and also lower the $^{12}$C/$^{13}$C ratio \citep{nol}  
(diagonal arrow in Fig.~\ref{fig2}). Any such  process would imply an anti-correlation 
between N and C isotopic ratios, but there is no evidence of this in N-type stars or in MS grains. Furthermore, 
the existence of grains with N isotopic ratios similar and/or lower than solar is not explained by stellar 
nucleosynthesis or galactic chemical evolution. This has been ascribed to isotopic fractionation or 
terrestrial contamination \citep{jed,ada} but now we have shown that some C(N) stars also lay on that region 
(although, being lower limits, their uncertainty prevents us 
from conclusive statements). The above situation, in any case, challenges our current understanding of stellar evolution. 

Finally, SC-type carbon stars are rare ($\lesssim 1\%$ of AGB carbon stars), 
indicating very short evolutionary times or uncommon evolutionary paths. With 
a C/O ratio very close to unity, O-rich and C-rich grain formation still relies 
on poorly-known chemical kinetic processes \citep{codust}. In fact, SC 
stars show little evidence of dust; and the solids that form include relatively 
uncommon species, like troilite (FeS) \citep{hon}. They may represent 
a short transition from C/O$\lesssim1$ to C/O$\gtrsim 1$ compositions;
but then N and C isotopic ratios should be close to those of N-type
stars and s-elements should always be enhanced. Instead, these stars are all 
$^{15}$N-rich (with $^{14}$N/$^{15}$N$ \lesssim 1000$), 
independently of their C isotope ratio; and s-element enhancements  exist in some, but not all 
of them. While the composition of these stars is consistent with both the MS and A+B $^{15}$N-rich grain
groups, their origin is a mystery. It was suggested that they
are massive ($\gtrsim 4$ M$_{\odot}$) O-rich AGB stars, forming
C-rich envelopes only for a short time due to an efficient
dredge-up \citep{fro}. The suspected larger masses would explain the
extreme Li enhancements observed in some of them through HBB and the
\citet{cam} mechanism. However, if HBB were sufficiently
active to produce a lot of Li, very low F, C/O and $^{12}$C/$^{13}$C
ratios and very large ($\gtrsim 10^4)$  $^{14}$N/$^{15}$N ratios would 
result, all clearly at odds with observations (Fig.~\ref{fig2}). 
In general, therefore, the chemical pattern of SC stars, including 
N isotopes, although in line with that of various presolar grains, 
cannot be explained by standard stellar evolution. Note that 
nuclear rate uncertainties in CNO cycling are {\bf not 
large enough} to account for 
the peculiar isotopic abundances we measured. Perhaps some of these stars 
formed with initial $^{14}$N/$^{15}$N ratios dispersed over a huge range, 
but physical paths leading to such a scenario are not known. Moreover, 
determinations of $^{14}$N/$^{15}$N in the local interstellar medium (ISM) yield values
for this ratio (290$\pm$40) close to the terrestrial and solar ones,
with a small gradient in distance moving away from the Galactic center
(21.1$\pm5.2$ kpc$^{-1} + 123.8\pm37.1$) \citep{ada}. Although such
measurements in the ISM are difficult, there seems to be no space
for a wide dispersion of values. So far, tentative explanations of the huge 
range of N ratios in SiC grains assumed contamination either from terrestrial N or from
cosmic-ray spallation (that should correlate with grain size and meteoritic age, respectively: \citet{jed}). 
Alternatively, non-equilibrium 
chemistry in the ISM might trigger isotopic fractionation \citep{ada,bon}. 
However, all these suggestions are now in conflict with the evidence that 
anomalous nitrogen isotopic admixtures already existed in the parent C-rich 
red giants.

Summing up, our new data establish observationally, for the first time that, while
C(N) stars are parents of MS grains with high $^{14}$N/$^{15}$N ratios, J-type carbon stars 
might generate A+B grains and SC stars might be a source for the grains with low $^{14}$N/$^{15}$N ratios.
However, no known evolutionary scenario can explain all the whole resulting evidence. One might guess 
that some mixing/nucleosynthesis mechanism occurs during a
stellar merging, producing peculiar stars as a result \citep{zha}. However, this is a qualitative 
speculation and the underlying physics is still largely unexplored.

\acknowledgments
Results are mainly based on observations made with the Italian Telescopio
Nazionale Galileo (TNG) operated on the island of La Palma by the Fundaci\'on Galileo Galilei of
INAF (Istituto Nazionale di Astrofisica) at the Spanish Observatorio del Roque de los Muchachos
of the Instituto de Astrof\'\i sica de Canarias. We thank K. Eriksson and R. Colin for providing the C-rich
atmosphere models and laboratory $^{12}$C$^{15}$N line positions, respectively. This work has been partially supported 
by Spanish grants AYA2008-04211-C02-02 and AYA-2011-22460. S.C. and O.S. acknowledge funding 
from FIRB-MIUR 2008 (RBFR08549F-002) and from PRIN-INAF 2011 Multiple populations in Globular 
Clusters: their role in the Galaxy assembly. 

{\it Facilities:} \facility{TNG (INAF-IAC)}.

\clearpage


\begin{figure}
\figurenum{1}
\centering
\includegraphics[scale=0.75]{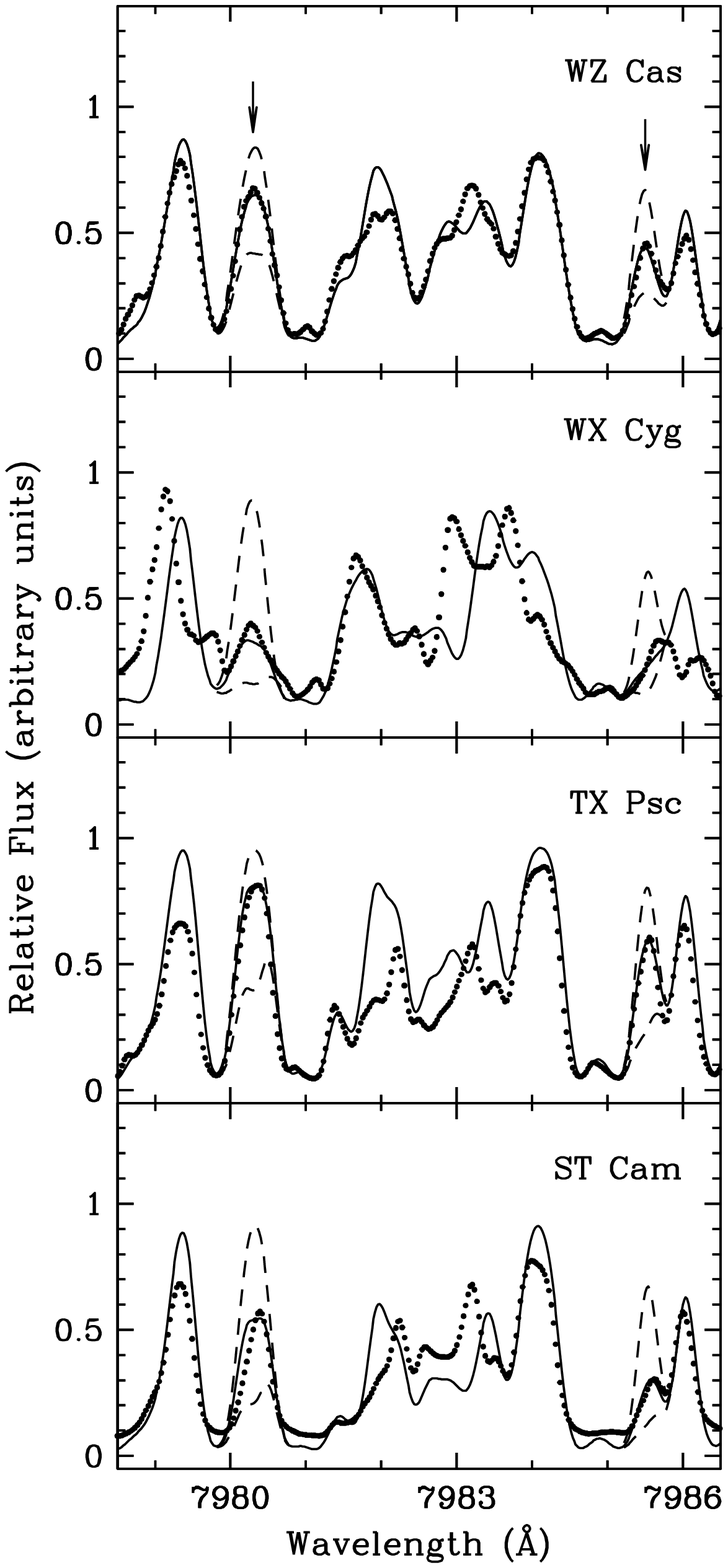}
\caption{\label{fig1} Comparison of observed (filled circles) with synthetic spectra (lines) for the
stars WZ Cas (SC-type), WX Cyg (J-type), TX Psc and ST Cam (N-type). The arrows in the
upper panel mark two of the $^{12}$C$^{15}$N features used to derive the $^{14}$N/$^{15}$N ratio.
For all stars the upper dashed line represents a synthetic spectrum calculation with no $^{15}$N; the
other lines represent spectra computed with $^{14}$N/$^{15}$N: 400 (continuous line), 100 (lower dashed 
line) for WZ Cas; 5, 1 for WX Cyg; 900, 100 for TX Psc; and 600, 100 for ST Cam, respectively.}
\end{figure}

\clearpage

\begin{figure}
\figurenum{2}
\centering
\includegraphics[scale=0.65]{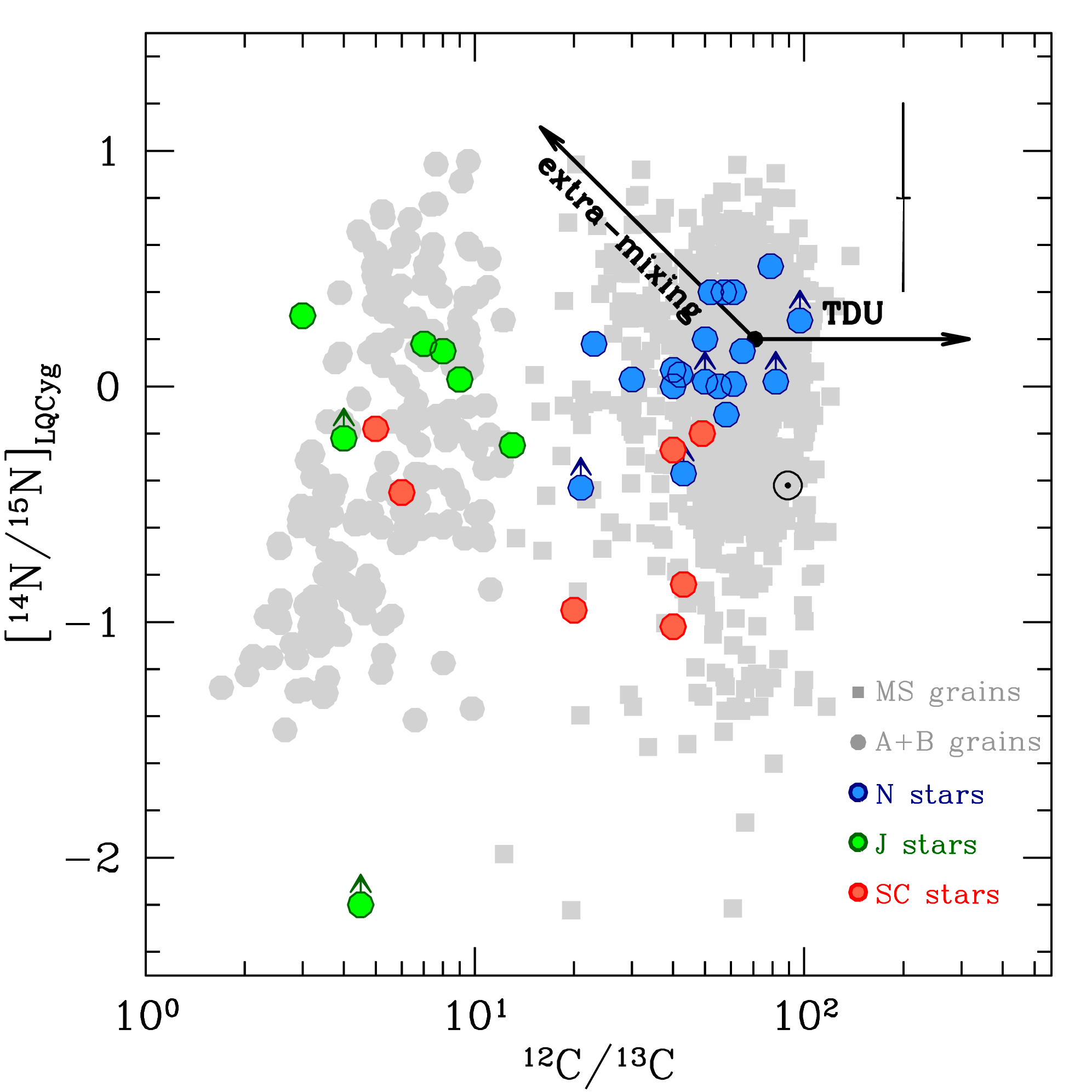}
\caption{\label{fig2} Colour circles: carbon and nitrogen isotopic ratios (relative to LQ Cyg)  derived for different C star 
spectral types. Grey symbols: isotopic ratios measured in MS and A+B SiC grains. A typical error 
bar on the stellar ratios is shown. Uncertainties in the SiC grains are smaller than the data points.
The black horizontal arrow indicates the theoretically expected behaviour of the C and N isotopic 
ratios during normal AGB evolution due to the TDU episodes. The diagonal arrow
instead represents the change in the isotopic ratios when an extra-mixing mechanism during the AGB is 
included (see text). The black dotted circle denotes the Solar System ratios.}
\end{figure}

\begin{deluxetable}{lcc}
\tablewidth{0pt}
\tablecolumns{3}
\tablecaption{Selected ${\rm ^{12}C^{15}N}$ lines \label{tab1}}
\tablehead{
\colhead{Wavelength} & \colhead{{$\chi$} } & \colhead{log $gf$}\\
\colhead{(\AA)} & \colhead{(eV)} & \colhead{} }
\startdata
7980.300 & 0.035 & -2.629 \\
7980.357 & 0.035 & -2.400 \\
7985.440 & 0.041 & -2.627 \\
7985.501 & 0.041 & -2.353 \\
7985.536 & 0.197 & -1.867 \\
8029.694 & 0.184 & -1.626 \\
8029.921 & 0.095 & -2.655 \\
8030.014 & 0.095 & -2.113 \\
8037.581 & 0.105 & -2.087 \\
8037.733 & 0.197 & -1.609 \\
8063.541 & 0.239 & -1.562 \\
\enddata
\tablecomments{
Several of these lines may contribute to the selected features in the analysis (see text).}
\end{deluxetable}

\clearpage

\begin{deluxetable}{lcccccc}
\tablewidth{0pt}
\tablecolumns{7}
\tablecaption{Nitrogen and carbon isotopic ratios \label{tab2}}
\tablehead{
\colhead{Star}                 &
\colhead{S/N}          & \colhead{$\rm[^{14}N/^{15}N]_{LQCyg}$}  &
\colhead{ ${\rm N_{lines}}$} & \colhead{$\rm ^{14}N/^{15}N$}& 
\colhead{$\rm^{12}C/^{13}C$} & \colhead{~~ref~~}} 
\startdata
N-type 		     & 	   & 		      &      &                &          &  \\ 
                     &     &                  &      &                &          & \\
AQ And 		     & 640 &	0.03$\pm$0.08&     3&  1230$\pm$260  &      30  & 2 \\ 
AW Cyg 		     & 410 &	$>$-0.43      &     2&  $>$750        &      21  & 4 \\ 
BL Ori$^{*}$ 	     & 620 &	0.40$\pm$0.20&     3&  3700$\pm$3000 &      57  & 1 \\ 
EL Aur 		     & 500 &	0.20$\pm$0.13&     4&  2300$\pm$1300 &      50  & 4 \\ 
LQ Cyg 		     & 290 &	0.0	      &     5&  1170$\pm$470  &      40  & 4 \\ 
NQ Gem 		     & 570 &	0.18$\pm$0.60&     2&  3700$\pm$3900 &      23  & 5 \\ 
ST Cam$^{*}$ 	     & 560 &	0.01$\pm$0.29&     4&  1300$\pm$1000 &      61  & 1 \\ 
SY Per 		     & 390 &	$>$-0.37      &     2&  $>$800        &      43  & 4 \\ 
TX Psc$^{*}$ 	     & 650 &	0.05$\pm$0.15&     3&  1040$\pm$150  &      43  & 1 \\ 
U Cam$^{*}$ 	  & 290 &     $>$0.28	   &	 2&  $>$2000  &      97  & 1 \\ 
UU Aur$^{*}$ 	     & 590 &	 $>$0.02      &     2&  $>$1000       &      52  & 1 \\ 
V460 Cyg$^{*}$	     & 560 &	0.40$\pm$0.29 &     2&  4600$\pm$2500 &     61  & 1 \\ 
V758 Mon 	     & 370 &	0.05$\pm$0.38&     4&  1600$\pm$1400 &      65  & 5 \\ 
V Aql$^{*}$	   & 580 &    $>$0.02	      &     2&  $>$1800       &      82  & 1 \\ 
W Cam 		     & 630 &	0.07$\pm$0.10&     3&  1300$\pm$200  &      40  & 6 \\ 
W Ori$^{*}$	     & 480 &	0.51$\pm$0.45&     3&  4300$\pm$2500 &      79  & 1 \\ 
X Cnc$^{*}$	     & 220 &	0.40$\pm$0.44&     3&  3300$\pm$1800 &      52  & 1 \\ 
Y Tau$^{*}$	     & 590 &   -0.12$\pm$0.25&     3&  880$\pm$190   &      58  & 1 \\ 
Z Psc$^{*}$	     & 570 &	0.00$\pm$0.45&     3&  1300$\pm$1100 &      55  & 1 \\ 
                     &  &    &     &    &      &  \\
J-type 		     &  &    &     &	&      &  \\ 
BM Gem 		     & 340 &	0.03$\pm$0.55&     2&  1330$\pm$800  &      9   & 3 \\ 
RX Peg 		     & 490 &	0.15$\pm$0.55&     2&  1800$\pm$1100 &      8   & 3 \\ 
UV Cam 		     & 440 &	 $>$-0.22     &     2&  $>$700        &      4   & 3 \\ 
V353 Cas 	     & 360 &	   0.18       &     1&  2400	      &      7   & 3 \\ 
V614 Mon 	     & 620 &	\nodata       &\nodata& \nodata      &       8   & 3 \\   
VX And$^{*}$	     & 650 &	 -0.25        &     1&  900	      &      13  & 1 \\ 
WX Cyg 		     & 230 &	   $>$-2.20   &     2&  $>$6	      &      4.5 & 3 \\ 
Y Cvn 		     & 500 &	0.30	      &     1&  3200	      &      3   & 3 \\
                     &     &                  &      &                &          &  \\
SC-type 	     &     &                  &      &                &	         &  \\ 
GP Ori$^{*}$	     & 240 &   -0.27$\pm$0.20&     5&  660$\pm$360   &      40  & 6 \\ 
RS Cyg 		     & 540 &   -1.02$\pm$0.15&     3&  105 $\pm$60   &      40  & 5 \\
RR Her 		     & 460 &   -0.84$\pm$0.29&     2&  220 $\pm$240  &      43  & 5 \\
RZ Peg$^{*}$ 	 & 260 &    \nodata   &  \nodata&  \nodata    &      12  & 6 \\   
UV Aur 		     & 260 &   -0.95$\pm$0.14&     3&  125$\pm$100   &      20  & 6 \\ 
VX Gem 		     & 220 &   -0.45$\pm$0.64&     2&  900$\pm$990   &      6   & 6 \\ 
WZ Cas$^{*}$	     & 380 &   -0.18$\pm$0.25&     3&  640$\pm$240   &      5   & 3 \\ 
BD +10 3764 	     & 630 &   -0.20$\pm$0.39&     2&  1100$\pm$1400 &      49  & 5 \\
\enddata
\tablecomments{$S/N$ is the signal to noise ratio achieved at $\sim$ 8000 {\AA}. ${N_{lines}}$ is
the number of $^{12}$C$^{15}$N lines used. The errors are the dispersion in N ratios when more than 
one $^{12}$C$^{15}$N line was used. For the stars marked with an asterisk, C and O
abundances were derived from the analysis of 2.2 $\mu$m spectra.} 
\tablerefs{Sources for $\rm{^{12}C/^{13}C}$: (1) Lambert et al. 1986; 
(2) Ohnaka \& Tsuji 1996; (3) Abia \& Isern 2000; (4) Abia et al. 2002; (5) Zamora et al. 2009; 
(6) derived in this work.}
\end{deluxetable}
\end{document}